\documentclass[a4paper,11pt]{article}
\usepackage[english]{babel}
\usepackage{pos}
\usepackage{amsmath}
\usepackage{mathtools}
\usepackage[compat=1.0.0]{tikz-feynman}
\usepackage{circuitikz}
\usepackage{dirtytalk}
\usepackage{multirow}
\usepackage{slashed}
\usepackage{changepage}

\def\glu{{\tilde{g}}}

\title{Antenna subtraction for final-state radiation at N$^3$LO}

\author*[a]{Petr Jakub\v{c}\'ik}

\affiliation[a]{Physik-Institut, Universität Zürich,\\
  Winterthurerstrasse 190, CH-8057 Zürich, Switzerland}
\emailAdd{petr.jakubcik@physik.uzh.ch}

\abstract{
I review some aspects of antenna subtraction at next-to-next-to-leading order (NNLO) in QCD and provide motivation for its extension to N$^3$LO. Next, I introduce the antenna functions required for the construction of infrared counterterms for final-state radiation at this order. Lastly, I describe the evaluation of the antenna functions and their phase-space integration, first presented in \cite{Jakubcik:2022zdi, Chen:2023fba, Chen:2023egx}, and I elaborate on their application to precision observables in jet production at lepton colliders.
}

\FullConference{Loops and Legs in Quantum Field Theory (LL2024)\\
 14-19, April, 2024\\
Wittenberg, Germany\\}

\begin{document}
\maketitle

\section{Introduction}
The past fifty years have seen the establishment of the Standard Model (SM) as the leading paradigm of particle physics. With the Large Hadron Collider (LHC), its last predicted constituent was discovered and the free parameters of the theory continue to be measured with ever-increasing accuracy in an attempt to test the model, break it, and ultimately advance to a new level of understanding. Besides, there is hope that the new physics which causes phenomena the SM cannot account for -- even if the associated energy scale is inaccessible -- might manifest as small deviations in precision observables in the High-Luminosity run of the LHC. On the theoretical side, this shift in how we make discoveries requires an upgrade from the leading-order calculations which helped identify the building blocks of the SM and the next-to-leading-order calculations which established it, to computational techniques which will test it at the ultimate, percent-level precision the LHC can reach.

For the case of fixed-order predictions in quantum chromodynamics (QCD), the current standard is next-to-next-to-leading order (NNLO). N$^3$LO computations are so far limited to processes with colour-singlet particles in the final state (and crossings thereof), when in fact \textit{jets} are the only observable QCD object and are most relevant for fitting the strong coupling or parton distribution functions (see~\cite{Caola:2022ayt} for a summary). 

Three main obstacles stand in the way of N$^3$LO calculations with jets. Firstly, the inherent increase in \textit{computational cost} and diminishing returns which are mitigated only by incremental technological improvements. Secondly, the availability of scattering \textit{matrix elements} which tie the definition of the theory to the building blocks of physical observables. Recent progress includes three-loop corrections to $2 \to 2$ scattering amplitudes with all on-shell \cite{Henn:2016jdu, Caola:2021izf, Caola:2021rqz, Caola:2022dfa} or one off-shell leg~\cite{Gehrmann:2023jyv}, and two-loop corrections to five-parton scattering~\cite{Agarwal:2023suw, DeLaurentis:2023izi, DeLaurentis:2023nss} or five-point with one off-shell leg~\cite{Abreu:2021asb}. Lastly, each matrix element contributing to a cross section at a fixed perturbative order needs to be integrated over the appropriate phase space, and their sum must be finite by the KLN theorem~\cite{Kinoshita:1962ur, Lee:1964is}. Nevertheless, the individual matrix elements are generically \textit{infrared (IR) divergent}: virtual (loop) corrections have explicit IR divergences in the dimensional regulator $\epsilon = (4-d)/2$, whereas real corrections diverge in regions of the phase space where particles become soft/collinear. Crucially, these infinities are not tangible until after the phase-space integration of the matrix element which cannot be performed analytically except in the simplest cases. 

\section{Infrared subtraction}
To overcome the last hurdle, one must identify the individual soft and collinear regions contributing to the divergence of the real(-virtual) radiation matrix element and use QCD factorisation theorems to describe each limit as a product of a simpler universal function which can be integrated analytically, and a Born-level matrix element. In other words, one needs to design a counterterm $\mathcal{S}$ to mimic the real correction $R$ in all unresolved regions and explicitly cancel the poles of the virtual contribution $V$ after integration over the unresolved partons, for example
\begin{align}
\sigma_{NLO} = \int \left[V  + \int \mathcal{S} d \phi_{1}\right] d \phi_{n} + \int \left[R- \mathcal{S}\right] d\phi_{n+1}\,,
\end{align}
where each integrand is now finite and suitable for numerical integration. This is the basis of \textit{subtraction} schemes. In general, these methods rely on 
understanding unresolved limits of matrix elements (see~\cite{Chen:2023egx} for a list of relevant calculations at N$^3$LO). As none of the established local subtraction schemes have yet been formulated and applied to processes with jets at N$^3$LO, in these proceedings I review the first steps taken within the framework of \textit{antenna subtraction}~\cite{Gehrmann-DeRidder:2005btv, Currie:2013vh}.

\subsection{Antenna subtraction}
Antenna subtraction is a promising candidate for extension to the next perturbative order due to its maturity: it is local (after angular averaging), it features analytic pole cancellation, it can handle hadrons in the initial state and jets in the final state and it boasts an efficient numerical implementation at NNLO. The main challenges in any subtraction scheme are the construction of the counterterm $\mathcal{S}$ which is process-dependent and its integration over the unresolved phase space. Schemes where the identification between parts of the counterterm and the unresolved limits is very tight must face the difficult integration of many soft/collinear functions. In antenna subtraction, the ingredients for the counterterms are simple matrix elements which are relatively easy to integrate but the complexity of the subtraction terms generally grows unfavourably with multiplicity, which has been addressed in recent publications~\cite{Gehrmann:2023dxm, Braun-White:2023sgd, Braun-White:2023zwd, Fox:2023bma}.

I briefly summarize how antenna subtraction works for final-state radiation at NNLO. The double-real matrix element $M^0_{n+2}$ can be described in the limit where one or two of the final state partons become soft/collinear as follows~\cite{Currie:2013vh}: 
\begin{align*}
\vcenter{\hbox{\resizebox{0.2\textwidth}{!}{\begin{circuitikz}
\tikzstyle{every node}=[font=\LARGE]
\draw [ fill={rgb,255:red,136; green,138; blue,133} , line width=1pt ] (.675,1.475) circle (.625cm);
\draw [line width=1pt, short] (-.875,2.35) -- (.25,1.925);
\draw [line width=1pt, short] (.25,1.025) -- (-.875,.6);
\draw [line width=1pt, dashed] (1.125,1.9) -- (2.125,2.6);
\draw [line width=1pt, dashed] (1.125,1.05) -- (2.125,.35);
\draw [line width=1pt, short, red] (1.275,1.675) -- (2.5,2.1);
\draw [line width=1pt, short, red] (2.5,.85) -- (1.275,1.275);
\draw [line width=1pt, dotted, blue] (2.625,1.6) -- (1.3,1.525);
\draw [line width=1pt, dotted, blue] (1.3,1.425) -- (2.625,1.35);
\node [font=\small] at (2.4,.23) {$n+2$};
\end{circuitikz}}}}
\to
\underbrace{
\left(
\frac{
\vcenter{\hbox{\resizebox{0.1\textwidth}{!}{\begin{circuitikz}
\tikzstyle{every node}=[font=\LARGE]
\draw [ fill={rgb,255:red,136; green,138; blue,133} , line width=0.5pt ] (.5,-.4) circle (.25cm);
\draw [line width=0.5pt, dashed] (-.125,-.4) -- (.25,-.4);
\draw [line width=0.5pt, dotted, blue] (.75,-.4) -- (1.125,-.4);
\draw [line width=0.5pt, short, red] (1.125,-.15) -- (.725,-.275);
\draw [line width=0.5pt, short, red] (.725,-.525) -- (1.125,-.65);
\end{circuitikz}}}}
}
{
\vcenter{\hbox{\resizebox{0.1\textwidth}{!}{\begin{circuitikz}
\tikzstyle{every node}=[font=\LARGE]
\draw [ fill={rgb,255:red,136; green,138; blue,133} , line width=0.5pt ] (.5,-.4) circle (.25cm);
\draw [line width=0.5pt, dashed] (-.125,-.4) -- (.25,-.4);
\draw [line width=0.5pt, short, red] (1.125,-.15) -- (.725,-.275);
\draw [line width=0.5pt, short, red] (.725,-.525) -- (1.125,-.65);
\end{circuitikz}}}}
}
\right)
}_{\text{antenna function}\; X_3^0}
\cdot
\vcenter{\hbox{\resizebox{0.2\textwidth}{!}{\begin{circuitikz}
\tikzstyle{every node}=[font=\LARGE]
\draw [ fill={rgb,255:red,136; green,138; blue,133} , line width=1pt ] (.675,1.475) circle (.625cm);
\draw [line width=1pt, short] (-.875,2.35) -- (.25,1.925);
\draw [line width=1pt, short] (.25,1.025) -- (-.875,.6);
\draw [line width=1pt, dashed] (1.125,1.9) -- (2.125,2.6);
\draw [line width=1pt, dashed] (1.125,1.05) -- (2.125,.35);
\draw [line width=1pt, short, red] (1.275,1.675) -- (2.5,2.1);
\draw [line width=1pt, short, red] (2.5,.85) -- (1.275,1.275);
\draw [line width=1pt, dotted, blue] (1.3,1.475) -- (2.65,1.475);
\node [font=\small] at (2.4,.23) {$n+1$};
\end{circuitikz}}}}
+ \begin{bmatrix} X_4^0 \\ X_3^0 X_3^0 \end{bmatrix}
\vcenter{\hbox{\resizebox{0.2\textwidth}{!}{\begin{circuitikz}
\tikzstyle{every node}=[font=\LARGE]
\draw [ fill={rgb,255:red,136; green,138; blue,133} , line width=1pt ] (.675,1.475) circle (.625cm);
\draw [line width=1pt, short] (-.875,2.35) -- (.25,1.925);
\draw [line width=1pt, short] (.25,1.025) -- (-.875,.6);
\draw [line width=1pt, dashed] (1.125,1.9) -- (2.125,2.6);
\draw [line width=1pt, dashed] (1.125,1.05) -- (2.125,.35);
\draw [line width=1pt, short, red] (1.275,1.675) -- (2.5,2.1);
\draw [line width=1pt, short, red] (2.5,.85) -- (1.275,1.275);
\node [font=\small] at (2.4,.23) {$n$};
\end{circuitikz}}}}
\end{align*}
The first term depicts the NLO-like limit which factorizes into the \textit{unintegrated antenna function} with 3 partons and 0 loops $X_3^0$ and the reduced matrix element $\tilde{M}^0_{n+1}$. The letter $X$ is a placeholder for a function name $A,\dots, H$ which depends on the partonic species, see~\cite{Gehrmann-DeRidder:2005btv}. The double unresolved limit is either iterated ($X_3^0 X_3^0$) or described by the four-parton tree-level antenna function $X_4^0$ and the Born-level matrix  element $\tilde{M}^0_{n}$.

The $\epsilon$-poles of the real-virtual matrix element due to single-unresolved radiation are described by the \textit{integrated antenna function} $\mathcal{X}_3^0$ defined above and one-loop single unresolved radiation, split into a tree $\times$ loop and loop $\times$ tree structure,
\begin{align*}
\vcenter{\hbox{\resizebox{0.2\textwidth}{!}{\begin{circuitikz}
\tikzstyle{every node}=[font=\LARGE]
\draw [line width=1pt ] (.675,1.475) circle (.625cm);
\draw [line width=1pt, short] (-.875,2.35) -- (.25,1.925);
\draw [line width=1pt, short] (.25,1.025) -- (-.875,.6);
\draw [line width=1pt, dashed] (1.125,1.9) -- (2.125,2.6);
\draw [line width=1pt, dashed] (1.125,1.05) -- (2.125,.35);
\draw [line width=1pt, short, red] (1.275,1.675) -- (2.5,2.1);
\draw [line width=1pt, short, red] (2.5,.85) -- (1.275,1.275);
\draw [line width=1pt, dotted, blue] (1.35,1.5125) -- (2.625,1.5125);
\node [font=\small] at (2.4,.23) {$n+1$};
\end{circuitikz}}}}
\to
\mathcal{X}_3^0
\cdot
\vcenter{\hbox{\resizebox{0.2\textwidth}{!}{\begin{circuitikz}
\tikzstyle{every node}=[font=\LARGE]
\draw [ fill={rgb,255:red,136; green,138; blue,133} , line width=1pt ] (.675,1.475) circle (.625cm);
\draw [line width=1pt, short] (-.875,2.35) -- (.25,1.925);
\draw [line width=1pt, short] (.25,1.025) -- (-.875,.6);
\draw [line width=1pt, dashed] (1.125,1.9) -- (2.125,2.6);
\draw [line width=1pt, dashed] (1.125,1.05) -- (2.125,.35);
\draw [line width=1pt, short, red] (1.275,1.675) -- (2.5,2.1);
\draw [line width=1pt, short, red] (2.5,.85) -- (1.275,1.275);
\draw [line width=1pt, dotted, blue] (1.35,1.5125) -- (2.625,1.5125);
\node [font=\small] at (2.4,.23) {$n+1$};
\end{circuitikz}}}}
+ X_3^0 \cdot
\vcenter{\hbox{\resizebox{0.2\textwidth}{!}{\begin{circuitikz}
\tikzstyle{every node}=[font=\LARGE]
\draw [ line width=1pt ] (.675,1.475) circle (.625cm);
\draw [line width=1pt, short] (-.875,2.35) -- (.25,1.925);
\draw [line width=1pt, short] (.25,1.025) -- (-.875,.6);
\draw [line width=1pt, dashed] (1.125,1.9) -- (2.125,2.6);
\draw [line width=1pt, dashed] (1.125,1.05) -- (2.125,.35);
\draw [line width=1pt, short, red] (1.275,1.675) -- (2.5,2.1);
\draw [line width=1pt, short, red] (2.5,.85) -- (1.275,1.275);
\node [font=\small] at (2.4,.23) {$n$};
\end{circuitikz}}}}
+ X_3^1 \cdot
\vcenter{\hbox{\resizebox{0.2\textwidth}{!}{\begin{circuitikz}
\tikzstyle{every node}=[font=\LARGE]
\draw [fill={rgb,255:red,136; green,138; blue,133} , line width=1pt ] (.675,1.475) circle (.625cm);
\draw [line width=1pt, short] (-.875,2.35) -- (.25,1.925);
\draw [line width=1pt, short] (.25,1.025) -- (-.875,.6);
\draw [line width=1pt, dashed] (1.125,1.9) -- (2.125,2.6);
\draw [line width=1pt, dashed] (1.125,1.05) -- (2.125,.35);
\draw [line width=1pt, short, red] (1.275,1.675) -- (2.5,2.1);
\draw [line width=1pt, short, red] (2.5,.85) -- (1.275,1.275);
\node [font=\small] at (2.4,.23) {$n$};
\end{circuitikz}}}}
\end{align*}
where grey circles denote tree-level matrix elements, otherwise the number of loops is indicated. Finally the poles of the double-virtual correction are entirely cancelled by integration of the antenna functions used above multiplied by lower-loop matrix elements:
\begin{align*}
\vcenter{\hbox{\resizebox{0.2\textwidth}{!}{\begin{circuitikz}
\tikzstyle{every node}=[font=\LARGE]
\draw [line width=1pt ] (.675,1.475) circle (.625cm);
\draw [line width=1pt, short] (-.875,2.35) -- (.25,1.925);
\draw [line width=1pt, short] (.25,1.025) -- (-.875,.6);
\draw [line width=1pt, dashed] (1.125,1.9) -- (2.125,2.6);
\draw [line width=1pt, dashed] (1.125,1.05) -- (2.125,.35);
\draw [line width=1pt, short, red] (1.275,1.675) -- (2.5,2.1);
\draw [line width=1pt, short, red] (2.5,.85) -- (1.275,1.275);
\draw [line width=1pt, short] (0.675,0.85) -- (0.675,2.1);
\node [font=\small] at (2.4,.23) {$n+1$};
\end{circuitikz}}}}
\to
\mathcal{X}_3^0
\cdot
\vcenter{\hbox{\resizebox{0.2\textwidth}{!}{\begin{circuitikz}
\tikzstyle{every node}=[font=\LARGE]
\draw [line width=1pt ] (.675,1.475) circle (.625cm);
\draw [line width=1pt, short] (-.875,2.35) -- (.25,1.925);
\draw [line width=1pt, short] (.25,1.025) -- (-.875,.6);
\draw [line width=1pt, dashed] (1.125,1.9) -- (2.125,2.6);
\draw [line width=1pt, dashed] (1.125,1.05) -- (2.125,.35);
\draw [line width=1pt, short, red] (1.275,1.675) -- (2.5,2.1);
\draw [line width=1pt, short, red] (2.5,.85) -- (1.275,1.275);
\node [font=\small] at (2.4,.23) {$n$};
\end{circuitikz}}}}
+ \begin{bmatrix}
    \mathcal{X}_4^0 \\ \mathcal{X}_3^1 \\ \mathcal{X}_3^0 \otimes \mathcal{X}_3^0
\end{bmatrix} \cdot
\vcenter{\hbox{\resizebox{0.2\textwidth}{!}{\begin{circuitikz}
\tikzstyle{every node}=[font=\LARGE]
\draw [ fill={rgb,255:red,136; green,138; blue,133}, line width=1pt ] (.675,1.475) circle (.625cm);
\draw [line width=1pt, short] (-.875,2.35) -- (.25,1.925);
\draw [line width=1pt, short] (.25,1.025) -- (-.875,.6);
\draw [line width=1pt, dashed] (1.125,1.9) -- (2.125,2.6);
\draw [line width=1pt, dashed] (1.125,1.05) -- (2.125,.35);
\draw [line width=1pt, short, red] (1.275,1.675) -- (2.5,2.1);
\draw [line width=1pt, short, red] (2.5,.85) -- (1.275,1.275);
\node [font=\small] at (2.4,.23) {$n$};
\end{circuitikz}}}}
\end{align*}
 An analogous sketch of the expected structure of the subtraction terms at N$^3$LO was presented in Eq.s (8)-(11) of~\cite{Marcoli:2023xtc}. In practice, one will start by enumerating the unresolved limits of the real radiation matrix element at N$^k$LO and proceed by partonic channel and colour factor (and colour ordering). First the relatively few $k$-fold unresolved limits are covered with simple combinations of the most divergent antenna functions. Since this intermediate counterterm also diverges in the many double- and single-unresolved configurations considered later, one has to avoid over-subtraction and the number of terms increases, which poses a computational but not conceptual obstacle.

\section{Antenna functions in final-final kinematics}
\subsection{Definition}
I turn my attention to the constituents of the subtraction terms. Antenna functions are the simplest matrix elements which capture the IR poles of all possible QCD radiation from a pair of \say{hard} partons. In the kinematic configuration where both radiators are in the final state, they are corrections to the decays of colour-singlet currents attached to a pair of radiators via a non-QCD vertex. At N$^k$LO, $X_n^{k+2-n}$ is the squared matrix element for the decay into $n$ partons at $k+2-n$ loops (normalised by the Born amplitude), and $\mathcal{X}_n^{k+2-n}$ is the integration over the inclusive $n$-particle phase space. Tab. 1 summarises the matrix elements and underlying QFTs from which we extracted the antenna functions with a pair of quarks, gluons or a quark and a gluon as hard radiators.

The quark-antiquark antenna functions can be extracted from either $\gamma^* \to q\bar{q}$ or $H\to b\bar{b}$ decay (with a Yukawa coupling but vanishing quark mass). The two processes deviate beyond the two leading $\epsilon$ poles due to the different Feynman rule in the coupling of the external current to quarks. Besides, the absence of $\gamma^\mu$ makes the configuration where the Higgs attaches to a closed quark loop (singlet) vanishing, unlike for the photon. Gluon-gluon antennae can be derived from corrections to the decay $H\to gg$ in the large top mass limit.

The quark-gluon case is pathological: a colour-singlet particle cannot decay into a $SU(3)$-fundamental and $SU(3)$-adjoint; a boson (like $\gamma, H$ above) cannot decay into a boson and a fermion. If we want to retain the correct spin for the hard radiators, the decaying particle has to be fermionic, and the products both $SU(3)$-adjoints. A theory in which this can be realised is the minimally-supersymmetric extension of the SM (MSSM) where a neutralino can decay via a heavy squark loop to a gluino (adjoint Majorana fermion) and a gluon. It suffices to work in a low-energy EFT~\cite{Haber:1988px} where the neutralino couples directly to the hard radiators and let $N_F$ flavours of quarks and $N_\glu$ flavours of gluinos propagate in the diagrams. The particulars of the non-QCD coupling are immaterial, the QCD radiation pattern between a gluon and a gluino emerges as we set $N_\glu = 0$, and relates to that of a gluon and a quark simply via the adjustment of colour factors, see Sec. 4 in~\cite{Chen:2023egx}.
\begin{table}[h!]
\begin{adjustwidth}{-0.25cm}{}
\bgroup
\def\arraystretch{1.03}%
\begin{tabular}{|c|c|c|}
\hline
\textbf{quark-antiquark}     & \textbf{gluon-gluon}         & \textbf{quark-gluon}         \\ \hline\hline
  Computed in:~\cite{Gehrmann-DeRidder:2004ttg},~\cite{Jakubcik:2022zdi} (\cite{Chen:2023fba}) & ~\cite{Gehrmann-DeRidder:2005alt},~\cite{Chen:2023fba}  & ~\cite{Gehrmann-DeRidder:2005svg},~\cite{Chen:2023egx}  \\ \hline
 $A,B,C$             & $G,H$               & $D,E,F,J,K$             \\ \hline
$q\bar{q}(g)$       & $gg(g)$       & $\glu g $       \\
$ggg$      & $q\bar{q}g$       &    $\glu gg(g);\,\glu q\bar{q}(g);\,\glu\glu'\glu'(g);\,\glu\glu\glu(g)$  \\
$q\bar{q}gg(g)$       & $gggg(g)$       &   $\glu g g g g;\ \glu q \bar{q} g g;\ \glu q \bar{q} q' \bar{q}'$     \\
$q\bar{q}q\bar{q}(g)$       & $ggq\bar{q}(g)$       &     $\glu q \bar{q} q \bar{q};\ \glu \glu' \glu' g g;\ \glu \glu \glu g g$   \\
$q\bar{q}q'\bar{q}'(g)$       &  $q\bar{q}q\bar{q}(g)$      &   $\glu \glu' \glu' q \bar{q};\ \glu \glu \glu q \bar{q};\ \glu \glu' \glu' \glu'' \glu''$     \\
       &   $q\bar{q}q'\bar{q}'(g)$   &   $\glu \glu' \glu' \glu' \glu';\ \glu \glu \glu \glu' \glu';\ \glu \glu \glu \glu \glu$  \\ \hline
 \rule{0pt}{20ex}    
\begin{tikzpicture}[thick, scale=0.65]

\begin{feynman}
\vertex (0) at (0,-3);
\vertex (1) at (2,-3);
\vertex (2u) at (3.03,-2.4);
\vertex (2d) at (3.03,-3.6);

\vertex [label=right:$\bar{q}$] (3u) at (6,-0.7);
\vertex [label=right:$q$] (3d) at (6,-5.3);
\vertex (cu) at (3.97,-2.5);
\vertex (cd) at (3.97,-3.5);
\vertex (c1) at (5.05,-2.5);
\vertex (c2) at (5.2,-3);
\vertex (c3) at (5.05,-3.5);
\vertex (e1) at (6.5,-2);
\vertex (e2) at (6.5,-3);
\vertex (e3) at (6.5,-4);

\node[blob,shape=ellipse,minimum height=0.95cm,minimum width=0.95cm] (blob) at (4.5, -3);
\node [fill=white,inner sep=1.5pt,rounded corners=2pt] at (4.5,-3) {\footnotesize QCD};
\diagram* {
(0) -- [photon, edge label = $\gamma^*\,(H)$ ] (1),
(3u) -- [fermion] (1),
(1) -- [ fermion] (3d),
(2u) -- [gluon ] (cu),
(cd) -- [gluon ] (2d),
(c1) -- [dotted ] (e1),
(c2) -- [dotted ] (e2),
(c3) -- [ dotted] (e3),
};
\end{feynman}
\end{tikzpicture}&
\begin{tikzpicture}[thick, scale=0.65]

\begin{feynman}
\vertex (0) at (0,-3);
\vertex (1) at (2,-3);
\vertex (2u) at (3.03,-2.4);
\vertex (2d) at (3.03,-3.6);

\vertex [label=right:$g$] (3u) at (6,-0.7);
\vertex [label=right:$g$] (3d) at (6,-5.3);
\vertex (cu) at (3.97,-2.5);
\vertex (cd) at (3.97,-3.5);
\vertex (c1) at (5.05,-2.5);
\vertex (c2) at (5.2,-3);
\vertex (c3) at (5.05,-3.5);
\vertex (e1) at (6.5,-2);
\vertex (e2) at (6.5,-3);
\vertex (e3) at (6.5,-4);

\node[blob,shape=ellipse,minimum height=0.95cm,minimum width=0.95cm] (blob) at (4.5, -3);
\node [fill=white,inner sep=1.5pt,rounded corners=2pt] at (4.5,-3) {\footnotesize QCD};
\diagram* {
(0) -- [scalar, edge label = $H$ ] (1),
(3u) -- [gluon] (1),
(1) -- [ gluon] (3d),
(2u) -- [dotted ] (cu),
(cd) -- [dotted] (2d),
(c1) -- [dotted ] (e1),
(c2) -- [dotted ] (e2),
(c3) -- [ dotted] (e3),
};
\end{feynman}
\end{tikzpicture}&
\begin{tikzpicture}[thick, scale=0.65]

\begin{feynman}
\vertex (0) at (0,-3);
\vertex (1) at (2,-3);
\vertex  (2u) at (3.03,-2.4);
\vertex  (2d) at (3.03,-3.6);

\vertex [label=right:$\glu$](3u) at (6,-0.7);
\vertex [label=right:$g$] (3d) at (6,-5.3);
\vertex (cu) at (3.97,-2.5);
\vertex (cd) at (3.97,-3.5);
\vertex (c1) at (5.05,-2.5);
\vertex (c2) at (5.2,-3);
\vertex (c3) at (5.05,-3.5);
\vertex (e1) at (6.5,-2) ;
\vertex (e2) at (6.5,-3);
\vertex (e3) at (6.5,-4) ;

\node[blob,shape=ellipse,minimum height=0.95cm,minimum width=0.95cm] (blob) at (4.5, -3);
\node [fill=white,inner sep=1.5pt,rounded corners=2pt] at (4.5,-3) {\footnotesize QCD};
\diagram* {
(1) -- [charged scalar, edge label = $\tilde{\chi}$ ] (0),
(3u) -- [fermion] (1),
(3u) -- [gluon] (1),
(1) -- [ gluon] (3d),
(2u) -- [gluon ] (cu),
(cd) -- [dotted ] (2d),
(c1) -- [dotted ] (e1),
(c2) -- [dotted ] (e2),
(c3) -- [ dotted] (e3),
};
\end{feynman}
\end{tikzpicture}\\ \hline
 $\mathcal{L}_{QCD} (+ y_b H \bar{\psi}\psi)$ & $\mathcal{L}_{QCD} - \frac{\lambda_{H}}{4} H G_a^{\mu\nu}G_{a, \mu\nu}$ & $\mathcal{L}_{QCD} + i \eta \overline{\psi}^a_{\glu} \sigma^{\mu\nu} 
\psi_{\tilde{\chi}} G_{\mu\nu}^a  + ({\rm h.c.})$ \\ \hline
QCD (+ Yukawa, $m_b = 0$) & QCD + HEFT & QCD + EFT of MSSM \\ \hline
 $\alpha_s$          & $\alpha_s$          & $\alpha_s$ ~\cite{Clavelli:1996pz}        \\
 ($y_{b}$~\cite{Gehrmann:2014vha})       & $\lambda_H$~\cite{Spiridonov:1988md}         & $\eta$~\cite{Chen:2023egx}     \\ \hline
\end{tabular}
\egroup
\end{adjustwidth}
\caption{Three types of matrix elements used to derive final-final antenna functions: reference for the NNLO and N$^3$LO calculations, conventional names for $X_n^{k+2-n}$, final states computed at N$^3$LO, representative Feynman diagrams, the relevant QFT and its Lagrangian, and the couplings which need to be renormalized. $\glu$~is a gluino and $\tilde{\chi}$ a neutralino, primes indicate different flavours of fermions. $H$ is the Higgs field,  $G_{\mu\nu}$ is the gluon field strength and $\psi_{\glu,\tilde{X}}$ the gluino and neutralino fields. $y_b, \lambda_H, \eta$ are the Yukawa $Hb\bar{b}$ coupling, the Higgs effective coupling to gluons in the large top mass limit, and the effective coupling of a neutralino to a gluino and gluon(s) via a low-energy EFT of the MSSM.}
\label{tab1}
\end{table}

\subsection{Computational method}
The obvious way to compute the unintegrated ($X_n^{k+2-n}$) and integrated ($\mathcal{X}_n^{k+2-n}$) antenna functions is to evaluate the amplitudes depicted in Tab. 1 and integrate inclusively each final state over the $n$-particle phase space. The complexity of a generic analytic phase-space integration is exactly the issue which subtraction schemes circumvent, and the power of antenna subtraction is that the integration of the IR counterterms can be rephrased using reverse unitarity~\cite{Anastasiou:2002yz,Anastasiou:2003gr}.

An arbitrary integral appearing in the N$^3$LO antenna function $\mathcal{X}_n^{5-n}$, featuring $n$ partons in the final state with momenta labelled by $p_\alpha$, $5-n$ loop momenta $k_\beta$, and a complete set of internal propagators with momenta $v_\gamma$ raised to integer powers, has the form
\begin{align}
    I_n^{5-n} = \underbrace{\int d \Pi_n}_{\text{p.s. integration}} \underbrace{(2\pi)^d \delta^{(d)}\left(\sum_{\alpha=1}^n p_\alpha\right)}_{\text{mom. conservation}} \underbrace{\int \prod_{\beta=1}^{5-n}\frac{d^d k_\beta}{(2\pi)^d}}_{\text{loop integration}} \underbrace{\prod_\gamma \frac{1}{(v_\gamma^{2})^{n_\gamma}}}_{\text{propagators}}\,.
\end{align}
Applying Cutkosky rules~\cite{Cutkosky:1960sp} to the $n$-particle phase space
\begin{align}
    d\Pi_n =& \prod_{\alpha=1}^n \frac{d^d p_\alpha}{(2\pi)^{d-1}}\delta(p_\alpha^2)\theta(p_\alpha^0)\\ 
-2\pi i \delta(p_\alpha^2)\theta(p_\alpha^0) \to&  \left(\frac{1}{p_{\alpha}^2+i0}-\frac{1}{p_\alpha^2-i0}\right) \equiv \frac{1}{\slashed{p}^2_\alpha}\,,
\end{align}
we turn the phase-space integration into a loop integration involving a set of $n$ \say{cut} propagators in a 4-loop 2-point integral with forward kinematics, effectively putting the two types of integration on equal footing,
\begin{align}
I_n^{5-n} = i^{n-1} \left(\int \prod_{\alpha=1}^n \frac{d^dp_\alpha}{(2\pi)^d}\frac{1}{\slashed{p}_\alpha^2}\right) \left(\int \prod_{\beta=1}^{5-n}\frac{d^d k_\beta}{(2\pi)^d}\prod_\gamma \frac{1}{(v_\gamma^{2})^{n_\gamma}} \right)(2\pi)^d  \delta^{(d)}\left(\sum_{\alpha=1}^n p_\alpha\right)\,.
\end{align}

We can therefore view the triple-real (RRR), double-real-virtual (RRV), RVV and VVV integrated antenna functions as the 5-, 4-, 3-, and 2- particle cuts of a 4-loop QCD correction to the $a = \gamma, H, \tilde{\chi}$ propagator. In this language, the sum of all the cuts (squared amplitudes with the $n$-parton final state $f_n$) relates to the imaginary part of the uncut amplitude via the optical theorem
\begin{align}
2\operatorname{Im}\left[\mathcal{M}(a\to a)\right] &= \sum_f \int d\Pi_f \mathcal{M}^{\dagger}(a\to f) \mathcal{M}(a\to f)\\
&= \sum_{f_5} \mathcal{X}_5^0 (f_5) + \sum_{f_4} \mathcal{X}_4^1 (f_4) + \sum_{f_3} \mathcal{X}_3^2 (f_3) + \sum_{f_2} \mathcal{X}_2^3 (f_2)\,.
\end{align}
This quantity is equal to the total decay width of $a$ into partons (jets), known as the $R$-ratio
\begin{align}
R = \frac{\sigma(\gamma^* \to \text{partons})}{\sigma(\gamma^* \to q\bar{q})}\,,
\end{align}
as an example for the photon. With these constraints in mind, the computation of the integrated antenna functions is best organised as follows:
\begin{enumerate}
    \item \textbf{UV poles of self-energies} \\
    Consider $\mathcal{O}(\alpha_s^3)$ corrections to the photon, Higgs and neutralino propagators. Using the dedicated programme FORCER~\cite{Ueda:2016yjm}, they can be reduced to a basis of known master integrals~\cite{Gituliar:2018bcr}. The $\epsilon$ poles of the imaginary part of the resulting expressions are entirely UV and with the knowledge of the lower loop orders, they determine the multiplicative \textit{renormalization of the relevant couplings} (see last row of Tab. 1).
    \item \textbf{Finite parts of self-energies} \\
    The finite remainders of the imaginary part of the renormalized self- energies are the $R$-ratios, which particularly for the neutralino is a new result beyond NLO, and they provide a \textit{check on the sum of all integrated antenna functions} of a given type (a column in Tab. 1).
    \item \textbf{All physical cuts of self-energies} \\
    Finally, one can place two, three, four or five cuts in all physical ways on the self-energy diagrams and tag contributions with different numbers of closed loops either side of the cut, different particle species on cut, and colour factors. The evaluation of the diagrams and reduction to master integrals was described in~\cite{Jakubcik:2022zdi}. The IR poles of these matrix elements which relate to the \textit{integrated antenna functions} are extracted with the help of the UV renormalization constants obtained above and the totals are checked for pole cancellation and the finite remainder against the $R$-ratios.
\end{enumerate}

\section{Results}
The ultimate result of the calculation is a \say{dictionary} of simple decay matrix elements (separated by partonic final state, loop configuration, colour layer and colour ordering), on one side integrated only over loops, and on the other side also over the phase space. The unintegrated tree-level $1\to5$, one-loop $1\to4$ and two-loop $1\to3$ matrix elements are well-known except in the neutralino case, and their unresolved limits will be compiled in a future publication. A better understanding of how contributions from individual unresolved limits cancel between the various real/virtual layers on the integrated level (between the antenna functions in the pole cancellation check and hence in future subtraction terms) is vital for simplifying and interpreting the scheme. Even though the integrated antenna functions have trivial kinematic dependence and are just $\epsilon$-expansions of transcendental numbers with rational coefficients, several patterns emerge.

\subsection{Structural observations}
\begin{itemize}
    \item \textbf{Triple-virtual matrix elements (vertex form factors)}\\
    We compared to existing calculations of the $\gamma^*\to q\bar{q}$, $H\to gg$~\cite{Gehrmann:2010ue} and $H\to b\bar{b}$~\cite{Gehrmann:2014vha} form factors which differ from the $\mathcal{X}_2^{3}$ only by a trivial phase-space integration. Moreover, the poles of VVV matrix elements ($1\to2$ decays) admit a description using universal constants (cusp and soft-collinear dimensions) and lower-loop-order matrix elements with Born kinematics~\cite{Becher:2009cu, Magnea:2018hab}. Consequently, the knowledge of the gluon-gluino dipole radiation for example determines the gluino contributions to the cusp, quark and gluon anomalous dimensions and the gluino collinear anomalous dimension up to $\mathcal{O}(\epsilon^3)$. Setting $N_F = 0$ and $N_\glu = 1$, we recover known results from $\mathcal{N}=1$ super-Yang-Mills theory~\cite{Grozin:2015kna}. 
    \item \textbf{Comparison between dipoles of different partons}\\
    Conversely for the infrared poles of real or mixed real-virtual radiation, no universal description is known beyond NLO. However, comparing the decays $\gamma^*\to q\bar{q}$ and $H\to b\bar{b}$, we see that the two deepest $\epsilon$ poles appearing in any colour layer are identical. This pattern extends to the coefficients of transcendental numbers in lower poles: for example in the $q\bar{q}q'\bar{q}g$ final state, the colour factor $N_FN$ starts at $\mathcal{O}(\epsilon^{-5})$ and the coefficient of $\zeta_3$ in the $\epsilon^{-2}$ pole ($=493/54$) or the coefficient of $\pi^4$ in the $\epsilon^{-1}$ pole ($=-3613/25920$) coincide between the two decays. A universal description of real radiation might receive distinct contributions from the two hard partons. Indeed, at least for the two deepest poles, we find that after replacing $C_F$ with $C_A$ to account for the difference between the quark and the gluino, schematically
    \begin{align}
        \mathit{Poles}(\tilde{\chi}\to g\glu) = \frac{\mathit{Poles}(\gamma^*\to q\bar{q})+\mathit{Poles}(H\to gg)}{2}\,,
    \end{align}
    see Eqs. (5.39)-(5.41) of~\cite{Chen:2023egx} for an exact formulation.
    \item \textbf{Notes on particular colour layers}\\
    From the inspection of colour factors, it is clear that multiple different integrated structures combine already in the deepest poles at N$^3$LO. For instance in the $q\bar{q}$ antenna functions, colour structures $C_F C_A$ and $C_A^2$ show up at $\mathcal{O}(\epsilon^{-6})$ which cannot simply come from the $I^{(1)}_{q\bar{q}}$ operator of~\cite{Catani:1998bh}. Often a colour factor only appears in two of the four layers, e.g. as a gluon-to-quarks splitting and the corresponding loop, such as in the singlet contribution to the four- and five-particle photon decays, and the cancellation is effectively NLO. Finally the most-subleading colour layers receive contributions from Abelian (photon-like) gluons and have a simple combinatorial explanation in terms of exponentiation of single emissions (see section 3.2 in~\cite{Chen:2023fba}). Sometimes this pattern of cancellation (though not the purely combinatorial explanation) extends also to more non-trivial factors like $N_F$ and $N_F N^{-2}$ of the $H\to q\bar{q}g(g)(g)$ final states.
\end{itemize}

\subsection{Phenomenological applications}
An immediate application of final-final antenna functions at N$^3$LO is the computation of the forward-backward asymmetry of bottom and charm quarks in $e^+ e^-$ annihilation in massless QCD. Denoting with $\sigma_{F, (B)}$ the cross section for the flavoured quark to be observed in the forward (backward) hemisphere as indicated by the jet axis of the flavoured quark jet, the observable is
\begin{align}
    A_{FB} = \frac{\sigma_F-\sigma_B}{\sigma_F+\sigma_B}\,.
\end{align}
The asymmetry is measured to 1\%-level accuracy even before the new generation of proposed lepton colliders (ILC, CLIC etc.) and particularly on the $Z$-resonance it is very sensitive to the effective weak mixing angle.

During the early approximate NNLO calculations~\cite{Altarelli:1992fs, Ravindran:1998jw}, an IR-safe definition of the hemispheres proved challenging, and the two results disagreed significantly. This shortcoming (due to $q\to qq\bar{q}$ splitting at NNLO) was addressed in~\cite{Catani:1999nf} and finally the full calculation was performed using antenna subtraction~\cite{Weinzierl:2006ij}. The NNLO corrections are already percent-level, making the asymmetry a precision observable, and the third-order corrections can now be computed including the two-loop~\cite{Gehrmann:2022vuk, Gehrmann:2023zpz} and three-loop~\cite{Gehrmann:2021ahy} matrix elements with axial-vector coupling.
Meanwhile the total 2-jet rate has already been computed at this order using unitarity and lower-order results in~\cite{Gehrmann-DeRidder:2008qsl}. Three-jet production, on the other hand, offers an array of event-shape observables highly sensitive e.g. to $\alpha_s$. The subtraction terms for this process are expected to be complicated and they require a proper definition of the quark-gluon antenna function from the neutralino decay matrix elements, as discussed in Sec. 5.2 of~\cite{Chen:2023egx}.

\section{Conclusion}
In this talk, I gave an overview of progress in extending antenna subtraction to N$^3$LO final-state radiation over the last two years. I argued that the definition of the scheme extends naturally from NNLO and described the evaluation and phase-space integration of the antenna functions necessary for building IR counterterms at this order. The presented ingredients are sufficient for the computation of the total rate and forward-backward asymmetry in 2-jet production at lepton colliders to $\mathcal{O}(\alpha_s^3)$.

For a full description of hadron collider processes, QCD radiation will also need to be described in initial-final and initial-initial kinematics. While the unintegrated antennas in the initial-final configuration are related by crossing, first steps in the evaluation of master integrals for the integrated functions were taken in~\cite{Fontana:2024olm}.

\section*{Acknowledgements}
I would like to thank my collaborators on this project Xuan Chen, Matteo Marcoli and Giovanni Stagnitto. I am grateful to Thomas Gehrmann for his advice and encouragement. This work was supported by the Swiss National Science Foundation (SNF) under contract 200020-204200 and by the European Research Council (ERC) under the European Union’s Horizon 2020 research and innovation programme grant agreement 101019620 (ERC Advanced Grant TOPUP).

\bibliographystyle{JHEP}
\bibliography{main}

\end{document}